\documentclass[seceq]{ptptex}

\usepackage{graphicx}

\newcommand{\ace}{\textit{ACE}}
\newcommand{\iras}{\textit{IRAS}}
\newcommand{\rosat}{\textit{ROSAT}}
\newcommand{\suzaku}{\textit{Suzaku}}
\newcommand{\wind}{\textit{WIND}}
\newcommand{\xmm}{\textit{XMM-Newton}}

\newcommand{\Ne}{\ensuremath{n_{\mathrm{e}}}}
\newcommand{\TLB}{\ensuremath{T_\mathrm{LB}}}
\newcommand{\logTLBsq}{\ensuremath{\log[\TLB/\mathrm{K}]}}
\newcommand{\OVII}{O~\textsc{vii}}
\newcommand{\OVIII}{O~\textsc{viii}}
\newcommand{\NeIX}{Ne~\textsc{ix}}

\newcommand{\cm}{\ensuremath{\mbox{cm}}}
\newcommand{\pc}{\ensuremath{\mbox{pc}}}
\newcommand{\s}{\ensuremath{\mbox{s}}}
\newcommand{\ph}{\ensuremath{\mbox{photons}}}
\newcommand{\counts}{\ensuremath{\mbox{counts}}}
\newcommand{\sr}{\ensuremath{\mbox{sr}}}
\newcommand{\erg}{\ensuremath{\mbox{erg}}}

\newcommand{\pcmsq}{\ensuremath{\cm^{-2}}}
\newcommand{\ps}{\ensuremath{\s^{-1}}}
\newcommand{\parcminsq}{\ensuremath{\mbox{arcmin}^{-2}}}
\newcommand{\psr}{\ensuremath{\sr^{-1}}}

\newcommand{\emismeas}{\ensuremath{\cm^{-6}} \pc}
\newcommand{\rassrate}{\counts\ \ps\ \parcminsq}
\newcommand{\surfbrig}{\erg\ \pcmsq\ \ps\ \psr}
\newcommand{\lineunit}{\ph\ \pcmsq\ \ps\ \psr}

\title{
\suzaku\ Observations of the Soft X-ray Background
}

\author{
David B. \textsc{Henley} \& Robin L. \textsc{Shelton}
}

\inst{
Dept. of Physics \& Astronomy, University of Georgia, Athens, GA 30602, USA
}

\abst{
We have analyzed a pair of \suzaku\ XIS1 spectra of the soft X-ray background, obtained by observing towards
and to the side of a nearby ($d = 230$~pc) absorbing filament in the southern Galactic hemisphere. We fit
multicomponent spectral models to the spectra in order to separate the foreground emission due to the Local
Bubble (LB) from the background emission due to the Galactic halo and unresolved AGN.

We obtain LB and halo parameters that are different from those obtained from our analysis of \xmm\ spectra
from these same directions. The LB temperature is lower ($\logTLBsq = 5.93$ versus 6.06), and the flux due
to the LB in the \suzaku\ band is an order of magnitude less than is expected from our \xmm\ analysis.
The halo components, meanwhile, are hotter than previously determined, implying our \suzaku\ spectra are
harder than our \xmm\ spectra.
}

\begin{document}

\maketitle

\section{Introduction}

Previous analyses have shown that several different components contribute to the soft X-ray background (SXRB),
including a hot bubble surrounding the solar neighborhood (the Local Bubble; LB), hot gas in the Galactic halo,
and unresolved AGN composing the extragalactic background. X-ray spectroscopy of the SXRB is
important as it enables us to determine the thermal properties, ionization
state, and chemical abundances of the X-ray-emitting plasma. This helps us constrain
models for the origin of the hot gas in the LB and halo, both
of which are poorly understood.

\begin{figure}
\centering
\begin{minipage}{0.45\linewidth}
\centering
\includegraphics[height=0.7\linewidth]{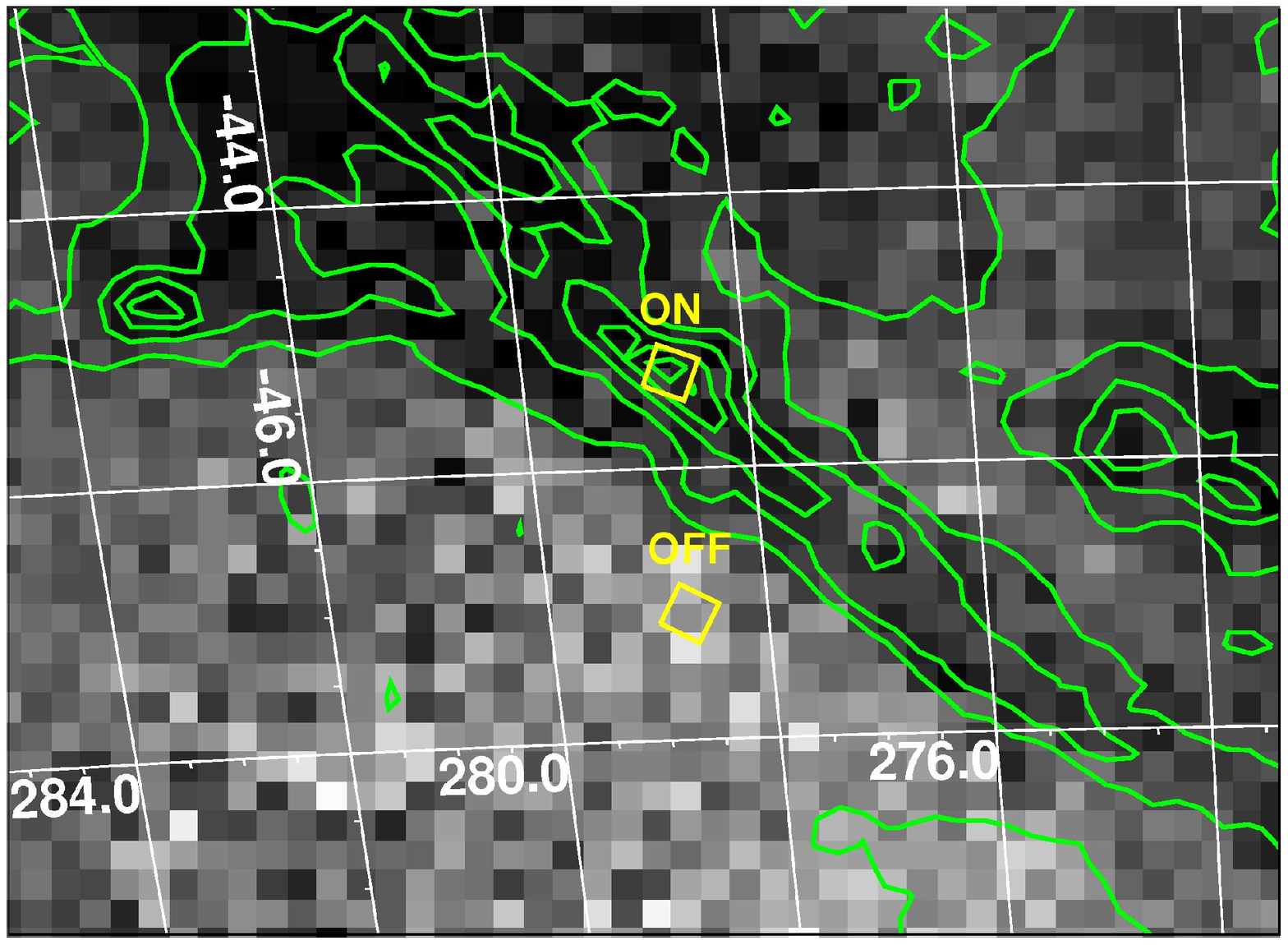}
\caption{
\rosat\ All-Sky Survey R12 intensity\cite{snowden97} (grayscale)
and \iras\ 100-micron intensity\cite{schlegel98} (contours), showing
the absorbing filament used for our observations. The yellow squares show our two
\suzaku\ pointing directions. \vspace{8.8em}
}
\label{fig:Shadow}
\end{minipage}
\hspace{0.05\linewidth}
\begin{minipage}{0.45\linewidth}
\centering
\includegraphics[height=0.7\linewidth]{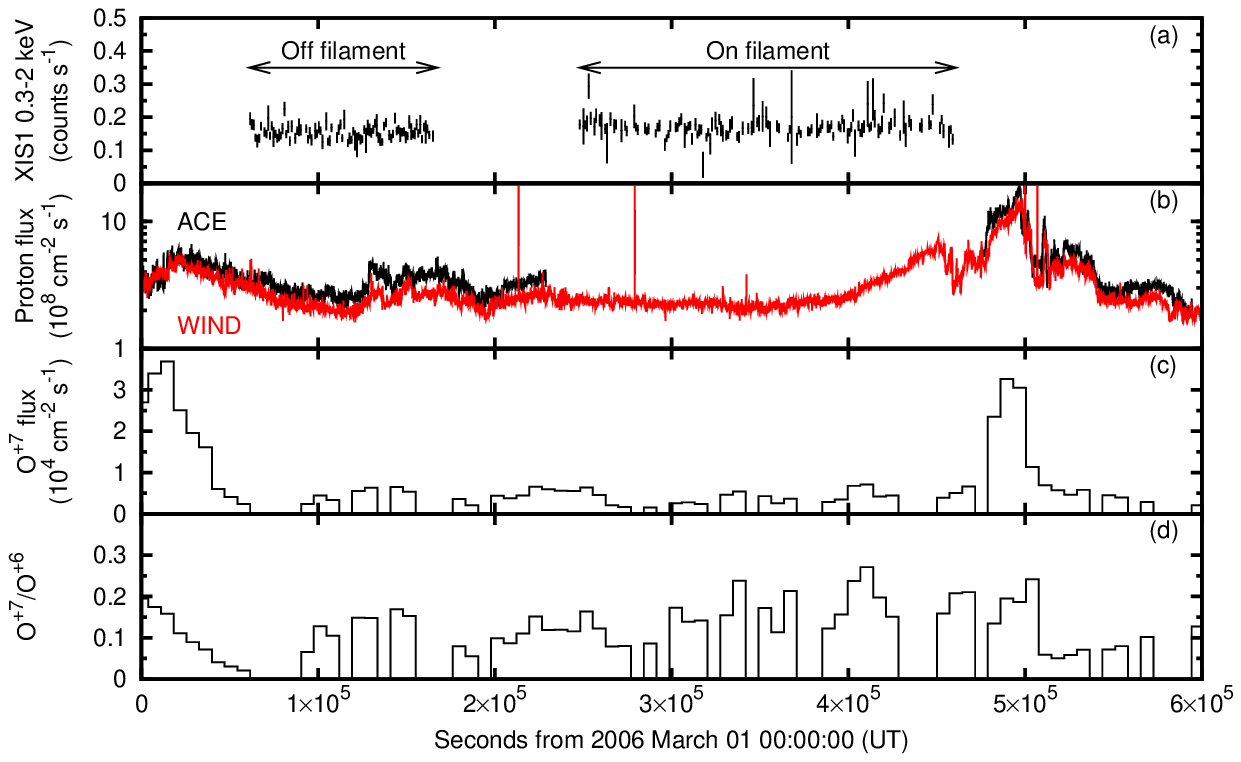}
\caption{
(a) \suzaku\ XIS1 0.3--2.0~keV count rate for the off- and on-filament observations, plotted in 1024-s bins against time since 2006 March 01.
    The particle background count rate has not been subtracted.
(b) Solar wind proton flux, from the \ace\ SWEPAM and \wind\ SWE experiments.
(c) Solar wind O$^{+7}$ flux, from the \ace\ SWICS experiment.
(d) Solar wind O$^{+7}$/O$^{+6}$ ion ratio, also from SWICS.
The solar wind data have not been shifted to allow for the travel time from the
spacecraft to the Earth ($\sim$3~ks).
}
\label{fig:ACE}
\end{minipage}
\end{figure}

In this paper we determine the spectra of the LB and the halo from a pair of \suzaku\ observations.
The \suzaku\ observations were obtained for pointings on and off a nearby absorbing filament in the
southern Galactic hemisphere ($b \approx -46^\circ$, $d = 230$~pc), which appears as a shadow in the soft X-ray
background (see Fig.~\ref{fig:Shadow}). Both observed spectra are expected to be of the form
\begin{displaymath}
\mbox{Local Bubble} + \mbox{Absorption} \times (\mbox{Halo} + \mbox{Extragalactic}).
\end{displaymath}
Since the absorbing column is different in the two pointing directions ($9.6 \times 10^{20}$~\pcmsq\ [on] versus
$1.9 \times 10^{20}$~\pcmsq\ [off]), fitting models simultaneously to both spectra enables one to disentangle
the LB and halo contributions.

\section{Spectral Analysis}

In our analysis, we concentrate only on data from the XIS1 detector, as this has the greatest sensitivity at low
energies. We simultaneously fit models consisting of LB, halo, and extragalactic components to our
on- and off-filament spectra. As indicated in \S1, the LB component is not subject to
any absorption, whereas the halo and extragalactic components are.
We modeled the Local Bubble plasma as an isothermal optically thin plasma, and the halo plasma as an optically
thin plasma having two temperature components. To model the plasmas' thermal emission, we used the APEC plasma
emission code.\cite{smith01} \ We modeled the extragalactic background's spectrum with a power-law.
To help constrain the model at low energies we included R12 data from the \rosat\ All-Sky Survey in the fit. To model
the \rosat\ emission we used the Raymond \& Smith code\cite{raymond77}, as APEC is not accurate below
$\sim$0.25~keV.\footnote{See http://cxc.harvard.edu/atomdb/issues\_caveats.html} \ The temperatures and emission
measures of the Raymond \& Smith components were tied to those of the corresponding APEC components.

\begin{figure}
\centerline{
	\includegraphics[width=0.47\linewidth]{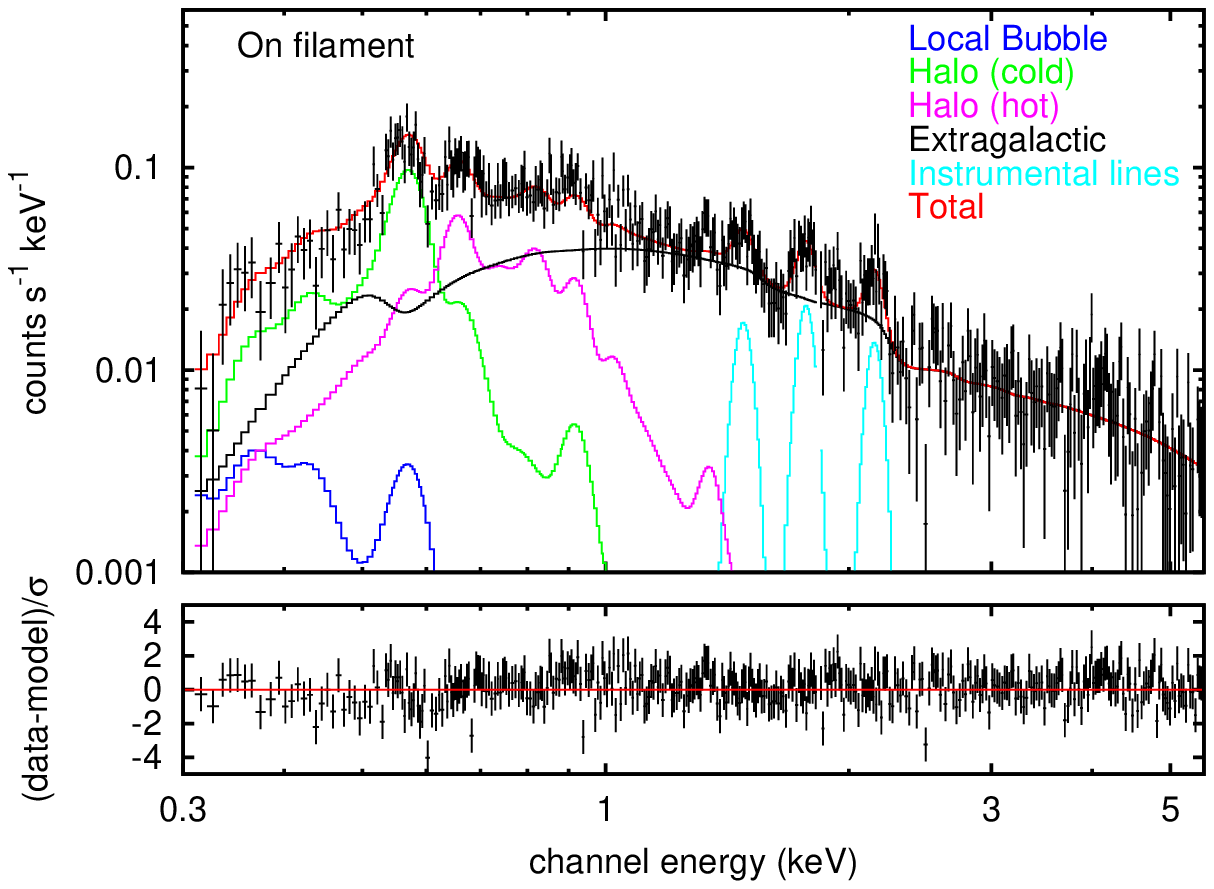}
	\hspace{0.04\linewidth}
	\includegraphics[width=0.47\linewidth]{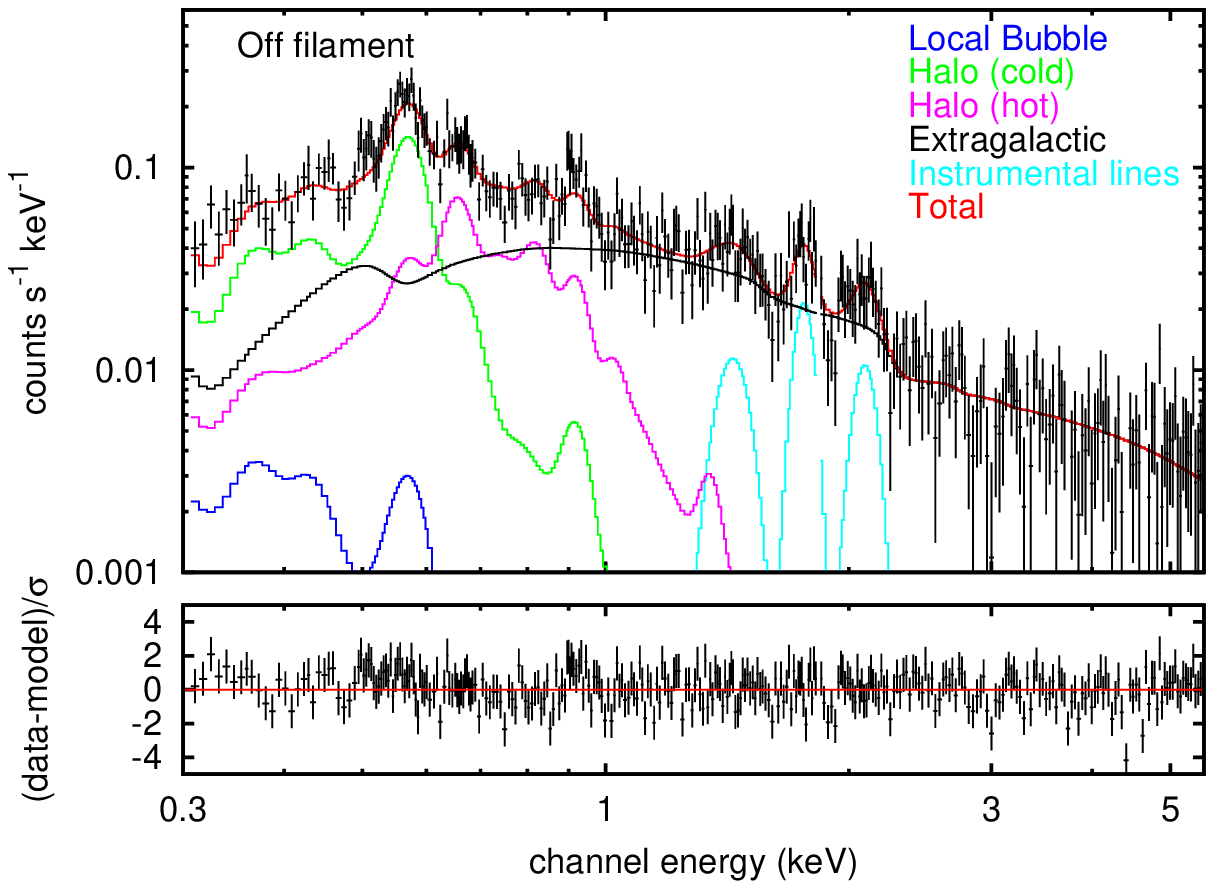}
}
\caption{
Observed on-filament (left) and off-filament (right) \suzaku\ spectra with our best-fitting model.
}
\label{fig:Spectra}
\end{figure}

Our observed spectra and best-fitting model are shown in Figure~\ref{fig:Spectra}. The best-fitting model parameters are shown
in Table~\ref{tab:Results} (model~A). For comparison, Table~\ref{tab:Results} also contains the results of our \xmm\ analysis for
these observation directions\cite{henley07} (model~B).

\begin{table}
\renewcommand\citeform[1]{#1}
\caption{
Our best fitting model parameters
}
\newcommand{\EM}{$\int \Ne^2 \mathrm{d}l$ (\emismeas)}
\begin{center}
\begin{tabular}{llcc}
\hline
                & 				& \suzaku\ $+$ \rosat\		& \textit{XMM} $+$ \rosat$^\mathrm{a}$	\\
		&				& Model A			& Model B				\\
\hline
Local Bubble    & $\log T$			& $5.93 \pm 0.04$		& $6.06^{+0.02}_{-0.04}$	\\
		& \EM				& 0.0043			& 0.018				\\
Halo (cool)	& $\log T$			& $6.15^{+0.02}_{-0.01}$	& $5.93^{+0.04}_{-0.03}$	\\
		& \EM				& 0.022				& 0.17				\\
Halo (hot)	& $\log T$ 			& $6.51 \pm 0.02$		& $6.43 \pm 0.02$		\\	
		& \EM				& 0.0057			& 0.011				\\
Exgal powerlaw	& Normalization$^\mathrm{b}$	& $10.5 \pm 0.2$		& 10.5$^\mathrm{c}$		\\
\hline
                & $\chi^2/\mbox{dof}$		& $700.28/692$			& $435.9/439$			\\
\hline
\multicolumn{4}{l}{$^\mathrm{a}$Ref.~\citen{henley07}} \\
\multicolumn{4}{l}{$^\mathrm{b}$keV~cm$^{-2}$~s$^{-1}$~sr$^{-1}$~keV$^{-1}$, assuming a photon index of 1.46\cite{chen97}$^)$} \\
\multicolumn{4}{l}{$^\mathrm{c}$Frozen; ref.~\citen{chen97}}

\end{tabular}
\end{center}
\label{tab:Results}
\end{table}


\section{Discussion}

The LB temperature we obtain from our \suzaku\ spectra is smaller than that obtained from our \xmm\ analysis,\cite{henley07} \
and also from analyses of \rosat\ All-Sky Survey data.\cite{kuntz00} \ The emission measure is also smaller than our \xmm-determined
value, though this is largely due to a difference between the spectral codes used.\cite{henley07} \ Both models predict
similar \rosat\ R12 count rates from the LB: $570 \times 10^{-6}$~\rassrate\ (model~A) versus
$587 \times 10^{-6}$~\rassrate\ (model~B), compared with the observed on-filament R12 count rate of
$(610 \pm 30) \times 10^{-6}$~\rassrate\ (the difference between the LB model count-rates and the observed count-rate can
be attributed to background [halo $+$ extragalactic] emission that has leaked through the filament).
The big difference between the models is in the \suzaku\ band: in model~A, the 0.3--0.7~keV
surface brightness of the LB ($1.2 \times 10^{-9}$~\surfbrig) is an order of magnitude smaller than in
model~B. The intensities of the LB \OVII\ emission at 0.57~keV in the two models are 0.13~\lineunit\ (model~A)
and 2.9~\lineunit\ (model~B). To put these results another way, the \suzaku\ spectra seem to be largely consistent with just
an absorbed halo $+$ extragalactic spectrum, without a significant foreground component. It is
currently unclear why our \suzaku\ spectra give different results from our \xmm\ spectra.

The halo parameters in model~A are also different from those in model~B. Both halo components are hotter in model~A than
in model~B, implying that the \suzaku\ spectra are harder than the \xmm\ spectra. Indeed, in the \suzaku\ spectra, the
Fe-L and \NeIX\ emission at 0.8--0.9~keV is brighter relative to the oxygen emission than in the \xmm\ spectra.
Also, the oxygen emission is distributed differently between the two halo components. In the \xmm\ model (model~B),
the hotter component is the source of $\sim$60\%\ of the halo \OVII\ emission and $\sim$95\%\ of the \OVIII\ emission,
whereas in the \suzaku\ model it is the source of only $\sim$20\%\ of the halo \OVII\ emission and $\sim$75\%\ of the
\OVIII\ emission. Again, it is not currently clear why the \suzaku\ spectra are different from the \xmm\ spectra.

One might wonder if solar wind charge exchange (SWCX) is affecting one or both of our spectra, which would render our modeling
method invalid.
We have examined solar wind data from \textit{ACE} and \textit{WIND}, to see if there are any changes in solar wind properties
between our observations. Various solar wind properties are plotted as functions of time in Figure~\ref{fig:ACE}, along with
the \suzaku\ XIS1 0.3--2.0~keV count rate. The solar wind proton flux
is fairly constant during the two observations, though it does start to rise towards the end of the on-filament observation.
However, there is not an associated rise in the soft X-ray count rate in this observations, nor does there seem to be
a change in the spectrum. Furthermore, the O$^{+7}$ flux and the O$^{+7}$/O$^{+6}$ ratio is similar for both observations, and
the values of these parameters are similar to what has been measured for other X-ray observations at times of low
SWCX contamination.\cite{snowden04, fujimoto06} \ It therefore seems unlikely that SWCX is significantly contaminating our spectra,
or, if it is, the SWCX emission should be the same for both observations, composing a fixed fraction of the foreground emission.


\section{Summary}

We have analyzed two \suzaku\ XIS1 spectra of the soft X-ray background, one obtained towards a nearby absorbing filament, and one
to the side of the filament. We simultaneously fit multicomponent spectral models to both spectra
in order to disentangle the Local Bubble (LB) and Galactic halo emission.

We obtain LB and halo parameters that are different from those obtained from \xmm\ spectra from these same directions.\cite{henley07} \
The LB component is $\sim$25\%\ cooler, and gives an order of magnitude less flux in the \suzaku\ band than is expected from our
\xmm\ analysis. The halo components, meanwhile, are both hotter, implying that our \suzaku\ spectra are harder than our \xmm\
spectra.

It is not currently clear why our \suzaku\ spectra give different results from our \xmm\ spectra. Inspection of contemporaneous
solar wind data implies that the discrepancy is not due to solar wind charge exchange contaminating our \suzaku\ spectra.
We are continuing to work on this problem.


\end{document}